\documentclass[prd,twocolumn,floatfix,preprintnumbers,showpacs]{revtex4}
\usepackage{graphicx}
\usepackage{dcolumn}
\usepackage{bm}
\begin{document}

\preprint{LAPTH-966/03, CERN-TH/2003-036, IFIC/03-07}

\title{Measuring the cosmological background of relativistic particles
with WMAP}

\author{Patrick Crotty$^1$, Julien Lesgourgues$^{2,1}$, Sergio Pastor$^3$}
\affiliation{$^1$ Laboratoire de Physique Th\'eorique LAPTH, B.P. 110, F-74941
Annecy-le-Vieux Cedex, France\\
$^2$ Theory Division, CERN, CH-1211 Geneva 23, Switzerland\\
$^3$ Instituto de F\'{\i}sica Corpuscular (CSIC-Universitat de
Val\`encia), Ed.\ Institutos de Investigaci\'on, Apdo.\ 22085,
E-46071 Valencia, Spain}

\date{May 7, 2003}

\begin{abstract}
We show that the first year results of the Wilkinson Microwave
Anisotropy Probe (WMAP) constrain very efficiently the energy density
in relativistic particles in the universe. We derive new bounds on
additional relativistic degrees of freedom expressed in terms of an
excess in the effective number of light neutrinos $\Delta N_{\rm
eff}$.  Within the flat $\Lambda$CDM scenario, the allowed range is
$\Delta N_{\rm eff} < 6$ (95\% confidence level) using WMAP data only,
or $ -2.6 < \Delta N_{\rm eff} < 4$ with the prior $H_0= 72 \pm 8$ km
s$^{-1}$ Mpc$^{-1}$. When other cosmic microwave background and large
scale structure experiments are taken into account, the window shrinks
to $ -1.6 < \Delta N_{\rm eff} < 3.8$. These results are in perfect
agreement with the bounds from primordial nucleosynthesis. Non-minimal
cosmological models with extra relativistic degrees of freedom are now
severely restricted.
\end{abstract}

\maketitle

\section{Introduction}

What is the matter budget of the universe? This fascinating and
central question in cosmology is currently being answered with
increasing precision, thanks to outstanding measurements of the
cosmic microwave background (CMB) anisotropies, correlated with the
study of large scale structure (LSS), of the primordial abundances
of light elements, and many other observables in the far universe.

One of the most intriguing issues is to determine the contributions of
matter and dark energy to the total energy density ($\Omega_m$ and
$\Omega_{\Lambda}$ in units of the critical density). Inflation
predicts $\Omega_m + \Omega_{\Lambda}=1$ but, so far, there is no
theoretical prediction on the value of each of these parameters. This
explains why in most cosmological parameter analyses they receive much
more attention than the radiation density $\Omega_r$, which is often
assumed to be well-known. However, the reference value of $\Omega_r$
relies on a strong theoretical prejudice: apart from the CMB photons,
the dominant relativistic backgound would consist of the three
families of neutrinos, whose temperature would be fixed with respect
to the CMB temperature by the standard picture of neutrino decoupling
prior to Big Bang Nucleosynthesis (BBN).

Measuring the energy density of radiation today is not easy, because
it is known to be three orders of magnitude below the critical
density.  The main constraints come either from the very early
universe, when radiation was the dominant source of energy, or from
the observation of cosmological perturbations, which imprint the time
of equality between matter and radiation. In particular, through BBN
models, the primordial abundances of light elements can be related to
$\Omega_r$, evaluated at the time when the mean energy in the universe
was of order $1~{\rm MeV}^4$ -- while the power spectrum of photon
anisotropies and of matter density carry a clear signature of the time
at which $\Omega_r$ became comparable to $\Omega_m$, at energies of
order $(0.1~{\rm eV})^4$. Until now, the BBN constraint on $\Omega_r$
was more stringent than the one from cosmological perturbations, by
approximately one order of magnitude. This leaves the door wide open
for various plausible assumptions concerning the radiation content of
the universe, which should not be necessarily the same during BBN and
at the time of matter/radiation equality.  For instance, a population
of non-relativistic particles may decay into relativistic ones,
enhancing the radiation energy density. Moreover, the standard BBN
scenario itself -- which is the simplest way to explain the formation
of light elements, but not the only one -- needs to be tested. In this
respect, the best would be to have some independent measurements of
the two free parameters of BBN: the baryon density $\Omega_b h^2$ and
the radiation density $\Omega_r h^2$ (where the reduced Hubble
constant is $h \equiv H_0/(100~{\rm km}~{\rm s}^{-1}~{\rm
Mpc}^{-1})$). The later controls the expansion rate in the early
universe. Some precise bounds on $\Omega_b h^2$ were already obtained
from recent CMB experiments, while the determination of $\Omega_r h^2$
was still quite loose, compared to BBN predictions.

The goal of this paper is to update this analysis and to show that the
outstanding data from the first year sky survey of the Wilkinson
Microwave Anisotropy Probe (WMAP) \cite{Bennett:2003bz,Spergel:2003cb}
gives better constraints, showing increasing evidence in favor of
standard BBN, and leaving very small room for extra relativistic
degrees of freedom beyond the three neutrino flavors.

\section{The effective number of relativistic neutrinos}

The energy density stored in relativistic species, $\rho_r$, is
customarily given in terms of the so-called {\it effective number of
relativistic neutrino species} $N_{\rm eff}$ (see \cite{Dolgov:2002wy}
for a review and references), through the relation
\begin{equation}
\rho_{\rm r} = \rho_\gamma + \rho_\nu + \rho_x =
\left[ 1 + \frac{7}{8} \left( \frac{4}{11}
\right)^{4/3} \, N_{\rm eff} \right] \, \rho_\gamma \,\,,
\label{neff}
\end{equation}
where $\rho_\gamma$ is the energy density of photons, whose value
today is known from the measurement of the CMB temperature. Eq.\
(\ref{neff}) can be also written as
\begin{equation}
N_{\rm eff} \equiv \left( \frac{\rho_r -\rho_\gamma}{\rho^0_\nu}\right)
\left(\frac{\rho^0_\gamma}{\rho_\gamma} \right)\, ,
\label{neff-def}
\end{equation}
where $\rho_\nu^0$ denotes the energy density of a single species of
massless neutrino with an equilibrium Fermi-Dirac distribution with
zero chemical potential, and $\rho^0_\gamma$ is the photon energy
density in the approximation of instantaneous neutrino decoupling. The
normalization of $N_{\rm eff}$ is such that it gives $N_{\rm eff} = 3$
in the standard case of three flavors of massless neutrinos, again in
the limit of instantaneous decoupling. In principle $N_{\rm eff}$
includes, in addition to the standard neutrinos, a potential
contribution $\rho_x$ from other relativistic relics such as majorons
or sterile neutrinos.

It turns out that even in the standard case of three neutrino flavors
the effective number of relativistic neutrino species is not exactly
3. The decoupling of neutrinos from the rest of the primordial plasma
occurs at a temperature of $2-3$ MeV, not far from temperatures of
order the electron mass at which electron--positron annihilations
transfer their entropy into photons, causing the well-known difference
between the temperatures of relic photons and relic neutrinos, $T/T_\nu
= \left( 11/4 \right)^{1/3}$ (see e.g.\ \cite{kt}). Accurate
calculations \cite{Hannestad:1995rs, Dolgov:1997mb, Esposito:2000hi}
have shown that neutrinos are still slightly interacting with $e^\pm$,
thus sharing a small part of the entropy release. This causes a
momentum dependent distortion in the neutrino spectra from the
equilibrium Fermi--Dirac behavior and a slightly smaller $T/T_\nu$
ratio. Both effects lead to a value of $N_{\rm eff} = 3.034$.  A
further, though smaller, effect on $T/T_\nu$ is induced by finite
temperature Quantum Electrodynamics (QED) corrections to the
electromagnetic plasma \cite{heckler,Lopez:1998vk}. A recent combined
study of the incomplete neutrino decoupling and QED corrections
concluded that the total effect corresponds to $N_{\rm eff} = 3.0395
\simeq 3.04$ \cite{Mangano:2001iu}. Therefore we define the extra
energy density in radiation form as
\begin{equation}
\Delta N_{\rm eff} \equiv N_{\rm eff}-3.04 \, \, .
\label{delta-neff}
\end{equation}
The standard value of $N_{\rm eff}$ corresponds to the case of
massless or very light neutrinos, i.e.\ those with masses much smaller
than 1 eV. More massive neutrinos affect the late avolution of the
universe in a way that can not be parametrized with a $\Delta N_{\rm
eff}$ . However, the recent evidences of flavor neutrino oscillations
in atmospheric and solar neutrinos, in particular after the recent
KamLAND data show that the neutrino masses are not large enough,
except in the case when the three mass eigenstates are degenerate
(see e.g.~\cite{Pakvasa:2003zv} for a recent review). We do not
consider such a case in the present paper, but assume that the
neutrino mass scheme is hierarchical, with the largest mass of order
$m_\nu \simeq \sqrt{\Delta m^2_{\rm atm}}\sim 0.05$ eV.

The value of $\Delta N_{\rm eff}$ is constrained at the BBN epoch from
the comparison of theoretical predictions and experimental data on the
primordial abundances of light elements. Typically, the BBN bounds are
of order $\Delta N_{\rm eff} < 0.4-1$
\cite{Lisi:1999ng,Esposito:2000hh,Kneller:2001cd,Cyburt:2001pq,Zentner:2001zr}.
Independent bounds on the radiation content of the universe can be
extracted from the analysis of the power spectrum of CMB anisotropies.
An enhanced contribution of relativistic particles delays the epoch of
matter-radiation equality, which in turn increases the early
integrated Sachs-Wolfe effect. Basically this leads to more power
around the scale of the first CMB peak. Previous analyses found weak
bounds on $\Delta N_{\rm eff}$
\cite{Jungman:1995bz,Hannestad:2000hc,Esposito:2000sv,
Hannestad:2001hn,Hansen:2001hi}, that can be significantly improved by
adding priors on the age of the universe or by including supernovae
and LSS data \cite{Hu:1998tk}. One of the most recent bounds on
$N_{\rm eff}$, from a combination of CMB and PSCz data
\cite{Hannestad:2001hn}, is $N_{\rm eff}=6^{+8}_{-4.5}$ (95\%
CL). Thus these bounds were not as restrictive as those from BBN.
However, the precise measurements of WMAP (and those of PLANCK in the
near future) are going to significantly improve the CMB constraint on
$\Delta N_{\rm eff}$, as shown in various forecast analyses (see for
instance \cite{Lopez:1998aq,Kinney:1999pd,Bowen:2001in}) and in the
calculations of this paper (see section \ref{results}).

Many extensions of the Standard Model of particle physics predict
additional relativistic degrees of freedom that will contribute to
$\Delta N_{\rm eff}$. There exist models with 4 neutrinos which
include an additional sterile neutrino in order to explain the third
experimental indication of neutrino oscillations (the LSND results).
It was shown in many studies (see for instance
\cite{DiBari:2001ua,Abazajian:2002bj}) that all four neutrino models,
both of 2+2 and 3+1 type, lead to a full thermalization of the sterile
neutrino flavor before BBN, and thus to $\Delta N_{\rm eff}\simeq 1$,
a value disfavored in the standard minimal model of BBN.  Moreover, in
these models there exists at least one neutrino state with mass of
order 1 eV.

It is also possible that the relativistic degrees of freedom at the
BBN and CMB epochs differ, for instance because of particle decays
which increase the photon temperature relative to the neutrino one
\cite{Kaplinghat:2000jj}. In some situations $\Delta N_{\rm eff}$ can
be effectively negative at BBN, such as the case of a distortion in
the $\nu_e$ or $\bar{\nu}_e$ spectra
\cite{Dolgov:1998st,Hansen:2000td}, or a very low reheating scenario
\cite{Giudice:2000ex}. 

A non-standard case that has been considered many times in the past
is the existence of relic neutrino asymmetries, namely when the
number of neutrinos and antineutrinos of the same flavor is
significantly different. These so-called degenerate neutrinos are
described by a dimensionless chemical potential
$\xi_\alpha=\mu_{\nu_\alpha}/T$, and it has been shown that the neutrino
energy density always increases for any value $\xi_\alpha\neq 0$
\begin{equation}
\Delta N_{\rm eff} = \sum_\alpha \left [ \frac{30}{7} \left
(\frac{\xi_\alpha}{\pi}\right )^2 + \frac{15}{7} \left
(\frac{\xi_\alpha}{\pi}\right )^4 \right ]
\label{delta-neff-xi}
\end{equation}
Interestingly, some combinations of pairs $(\xi_e,\xi_{\mu,\tau})$
could still produce the primordial abundances of light elements for a
larger baryon asymmetry, in the so-called degenerate BBN scenario
\cite{Kang}. At the same time, the weaker CMB bounds on $\xi_\nu$ are
flavor blind \cite{paper1,Kinney:1999pd}. However, it was recently
shown that for neutrino oscillation parameters in the regions favored
by atmospheric and solar neutrino data flavor equilibrium between all
active neutrino species is established well before the BBN epoch
\cite{Dolgov:2002ab,Wong:2002fa,Abazajian:2002qx}. Thus the stringent
BBN bounds on $\xi_e$ apply to all flavors, so that the contribution
of a potential relic neutrino asymmetry to $\Delta N_{\rm eff}$ is
limited to very low values.

\section{Method}

Using the {\tt cmbfast} code \cite{Seljak:1996is}, we computed the
cosmological perturbations (temperature and polarization anisotropies
$C_l^{(T,E,TE)}$, matter power spectrum $P(k)$) for a grid of models
with the following parameters: baryon density $\omega_b = \Omega_b
h^2$, cold dark matter density $\omega_{cdm} = \Omega_{cdm} h^2$,
hubble parameter $h$, scalar tilt $n_s$, optical depth to reionization
$\tau$, global normalization -- which is not discretized -- and of
course an additional contribution from relativistic particles $\Delta
N_{\rm eff}$. We include corrections to the CMB spectra from
gravitational lensing \cite{Zaldarriaga:1998ar}, as computed by {\tt
cmbfast}.  Apart from $\Delta N_{\rm eff}$, our set of parameters is
the simplest one used by the WMAP team in their parameter analysis
\cite{Spergel:2003cb}, and accounts very well for the first year WMAP
data. We restrict ourselves to a flat universe: since the curvature is
known to be small from the position of the first CMB peak, we adopt
the theoretical prejudice that the universe is exactly flat, as
predicted by inflation, rather than almost flat.  Therefore,
$\Omega_\Lambda$ is equal to $1 - (\omega_b +
\omega_{cdm})/h^2$. Allowing for a small curvature could alter our
results by a few percent.  We also neglect the possible contribution
of gravitational waves and a possible scale-dependent tilt.  A running
tilt, in favor of which the WMAP collaboration finds some marginal
evidence, would not change our predictions based on WMAP alone,
because the later does not constrain the primordial spectrum on a wide
enough range of scales. However, it could slightly alter our results
based on CMB and LSS data.

Our grid covers the following ranges: $0.019 < \omega_b < 0.028$,
$0.065 <\omega_{cdm}< 0.27 $, $0.5 < h < 0.9$, $0.8<n_s<1.28$,
$0<\tau<0.5$, $-3 < \Delta N_{\rm eff} < 5$.  We analyze it using an
interpolation and minimization routine developed at LAPTH. Our code
performs a multi-dimensional interpolation for each value of $C_l$ or
$P(k)$ in order to obtain the spectrum at any arbitrary point, and
then, computes the likelihood of the model. For WMAP, the likelihood
is calculated using the software kindly provided at the NASA web
site \cite{nasaweb}, and explained in \cite{Verde:2003ey}.  We will
also define a combined likelihood including the pre--WMAP CMB data
compilation by Wang et al. \cite{Wang:2002rt} (which is still useful
for constraining high multipoles), and the LSS data derived by Percival
et al.\ \cite{Percival:2001hw} (32 points on wavenumbers 
$k < 0.15~h \,$Mpc$^{-1}$) from the 2dF redshift survey
\cite{Hawkins:2002sg}. For these two data sets, we use window
functions and correlation matrices available from
\cite{tegmarkwww,2dFwww}. We constrain each free parameter using a
Bayesian approach: the 68\% (resp.\ 95\%) confidence limits are defined
as the values for which the marginalized likelihood drops by $\exp [
-(\chi_0^2 - 1)/2 ]$ (resp. $\exp [ -(\chi_0^2 - 4)/2 ]$), where
$\chi_0^2$ is the best chi-square value in the whole parameter space.

\begin{figure}
\includegraphics[angle=-90,width=.5\textwidth]{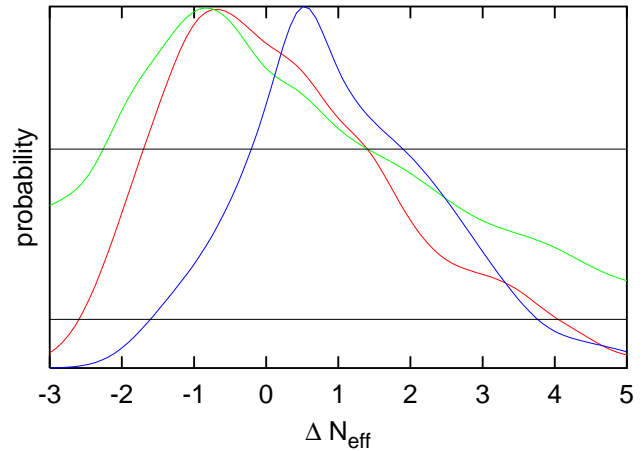}
\caption{\label{deltaN} The $\Delta N_{\rm eff}$ likelihood for WMAP +
weak $h$ prior (green), WMAP + strong $h$ prior (red), and the same
plus other CMB experiments and the 2dF redshift survey (blue). The
horizontal lines show the 68\% (resp.\ 95\%) confidence levels. The
step of $\Delta N_{\rm eff}$ in our grid of models is $0.5$.}
\end{figure}

\section{Results and discussion}
\label{results}

We start the analysis using only the first year WMAP temperature and
polarization data, plus a weak prior $0.5 < h < 0.9$, which is
implicit from the limitation of the grid. We checked that for $\Delta
N_{\rm eff}=0$, we find the same bounds as the WMAP collaboration,
with a minimal effective chi square $\chi^2_{\rm eff} = 1431.5$ for
1342 effective degrees of freedom (d.o.f.).  The best fit over our
whole grid has $(\Delta N_{\rm eff}, \omega_b, h, \omega_{cdm}, \tau,
n_s) = (-0.9, 0.024, 0.68, 0.10, 0.16, 0.97)$, and a $\chi^2$ value
which is not significantly lower ($\chi^2_{\rm eff} = 1431.1$ for 1341
d.o.f.), showing that a non--zero $\Delta N_{\rm eff}$ is {\it not}
required in order to improve the goodness of fit. At the 95\%
confidence level (CL), we can derive only an upper bound $\Delta
N_{\rm eff} < 6$, which is impressively smaller than the previous
bound from CMB only, $\Delta N_{\rm eff} < 14$
\cite{Hannestad:2001hn,Bowen:2001in}.  At the 1-$\sigma$ level, the
extra relativistic energy density is limited to the range $-2.2<
\Delta N_{\rm eff} < 1.2$, corresponding to a dispersion of 3.4. This is
in nice agreement with the prediction from \cite{Bowen:2001in}
that with the full WMAP data, one would reach a dispersion of 3.17.
The analysis reveals that the indetermination of this parameter is
caused mainly by a degeneracy with $h$, as shown in previous works.
When $\Delta N_{\rm eff}$ runs from -3 to 6, the best-fit value of
$h$ goes from 0.55 to 0.90 (while $\omega_{cdm}$ decreases,
maintaining an approximately constant value of the matter fraction
$\Omega_m= 0.31 \pm 0.03$). After $\Delta N_{\rm eff}=6$, the weak
prior $h<0.9$ is saturated, and the probability drops abruptly.

In order to remove the degeneracy, we add to the definition of the
effective $\chi^2$ a gaussian prior on $h$ derived from the HST Key
Project \cite{HST}, $h=0.72 \pm 0.08$ (at the 1-$\sigma$ level). This
prior is sufficient to reduce significantly the 95\% allowed window:
$-2.6 < \Delta N_{\rm eff} < 4.0$. The no neutrino case with $\Delta
N_{\rm eff}=-3$ is compatible with the data only beyond 99\% CL.  The
best-fit model has still a slightly negative $\Delta N_{\rm
eff}=-0.7$, but this is not statistically significant. The best
$\chi^2_{\rm eff}$ does not change very much (1431.1 for 1342 d.o.f.)
because WMAP alone is in remarkable agreement with the Key Project
value \cite{Spergel:2003cb}.

\begin{figure}[t]
\includegraphics[width=.5\textwidth]{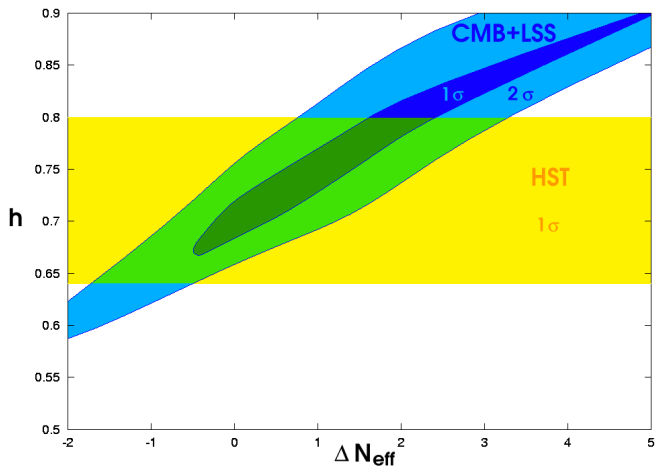}
\caption{\label{deltaN_h} The two-dimensional confidence limits
on ($\Delta N_{\rm eff}$, $h$), based on CMB and LSS data,
at the 1-$\sigma$ (dark blue) and
2-$\sigma$ (light blue) levels. The superimposed yellow stripe shows
the HST Key Project measurement of $h$ (1-$\sigma$ level).}
\end{figure}

\begin{table}[t]
\caption{The best fit values and 2-$\sigma$ (95\% CL) limits on $\Delta
N_{\rm eff}$ for the different data sets.}
\begin{tabular}{lc}
data set & $\Delta N_{\rm eff}$ \cr \hline \hline
& \cr
WMAP + weak $h$ prior & $-0.9^{+6.9}_{-2.1}$ \cr
& \cr
WMAP + strong $h$ prior & $-0.7^{+4.7}_{-1.9}$ \cr
& \cr
WMAP + other CMB + LSS + strong $h$ prior & $0.5^{+3.3}_{-2.1}$
\cr
& \cr
\hline
\end{tabular}
\label{table1}
\end{table}

Finally, when we include the pre-WMAP data compilation and the 2dF
redshift survey (i.e., information on the third and fourth CMB peak,
and on the scale of the turn--over in the matter power spectrum), the
window tends to shift a little bit towards positive values of $\Delta
N_{\rm eff}$. The best-fit model has $\Delta N_{\rm eff}=0.5$ and
$\chi^2_{\rm eff}=1492$ for 1396 d.o.f. The $95\%$ allowed window is
$-1.6 < \Delta N_{\rm eff} < 3.8$. Using all data, we find that the no
neutrino case is excluded at $99.9\%$ CL, which constitutes a clear
indication of the presence of relic background neutrinos, already
shown by pre-WMAP data \cite{Hannestad:2001hn}.

Note that throughout the analysis, we did not include any constraint
from supernovae data. This is because all good--fitting models
have naturally ($\Omega_m$, $\Omega_{\Lambda}$) very close to (0.3,
0.7).  So, adding a supernovae prior would be completely irrelevant -
unlike the $h$ prior, which plays a crucial role in removing the
degeneracy with $\Delta N_{\rm eff}$. In order to emphasize this last
point, we show in figure \ref{deltaN_h} the two-dimensional 1-$\sigma$
and 2-$\sigma$ confidence limits in ($\Delta N_{\rm eff}$, $h$) parameter
space. The contours are based only on the CMB and LSS data (the HST
Key Project result is just superimposed as a yellow stripe).  The
degeneracy between these two parameters clearly appears, showing that
any improvement in the direct determination of $h$ will be crucial for
closing the $\Delta N_{\rm eff}$ allowed window.  For instance, large
deviations from the standard value $\Delta N_{\rm eff}=0$ would be
excluded if $h$ were measured to be very close to $0.70$, while $h>0.75$
would bring strong evidence for extra relativistic species.

To summarize, we show in table \ref{table1} the best fit values and
2-$\sigma$ (95\% CL) limits on $\Delta N_{\rm eff}$ for the three
different data sets.  Our results show that the first year results of
WMAP significantly improve the bounds on an additional contribution to
the radiation density of the universe. In the near future, the updated
data from WMAP and the new results from the Sloan Digital Sky Survey
\cite{Loveday:2002ax} should allow for an even better determination,
providing a high-precision test of primordial nucleosynthesis.\\

{\bf Note added:} After the submission of this work, our results were
confirmed in refs.\ \cite{Pierpaoli:2003kw} (with a generalization of
the bounds to a non-flat Universe) and \cite{Hannestad:2003xv}.

\section*{Acknowledgments}
We are grateful to the WMAP collaboration for providing a
user-friendly access to their data, and in particular to Licia Verde
for her very useful comments. This work was supported by a
CICYT--IN2P3 agreement. SP was supported by the Spanish grant
BFM2002-00345 and a Marie Curie fellowship under contract
HPMFCT-2002-01831. SP thanks CERN for support during a visit when this
work was initiated.

\end{document}